\documentclass[aps,twocolumn,showpacs,preprintnumbers,amsmath,amssymb,prb,superscriptaddress]{revtex4-1}
\usepackage{amsmath}
\usepackage{graphicx}
\usepackage{dcolumn}
\usepackage{color}
\usepackage{float}

\newcommand{\be}{\begin{equation}}
\newcommand{\ee}{\end{equation}}
\newcommand{\bea}{\begin{eqnarray}}
\newcommand{\eea}{\end{eqnarray}}
\newcommand{\bs}{\begin{split}}
\newcommand{\bse}{\begin{subequations}}
\newcommand{\ese}{\end{subequations}}

\begin{document}
\title{Physical properties of noncentrosymmetric superconductor LaIrSi$_{3}$: A $\mu$SR study}

\author{V. K. Anand}
\altaffiliation{vivekkranand@gmail.com}
\affiliation{ISIS Facility, Rutherford Appleton Laboratory, Chilton, Didcot, Oxon, OX11 0QX, United Kingdom}
\affiliation{Helmholtz-Zentrum Berlin f\"{u}r Materialien und Energie, Hahn-Meitner Platz 1, D-14109 Berlin, Germany}
\author{D. Britz}
\affiliation {Highly Correlated Matter Research Group, Physics Department, University of Johannesburg, P.O. Box 524, Auckland Park 2006, South Africa}
\author{A. Bhattacharyya}
\affiliation{ISIS Facility, Rutherford Appleton Laboratory, Chilton, Didcot, Oxon, OX11 0QX, United Kingdom}
\affiliation {Highly Correlated Matter Research Group, Physics Department, University of Johannesburg, P.O. Box 524, Auckland Park 2006, South Africa}
\author{D. T. Adroja}
\altaffiliation{devashibhai.adroja@stfc.ac.uk}
\affiliation{ISIS Facility, Rutherford Appleton Laboratory, Chilton, Didcot, Oxon, OX11 0QX, United Kingdom}
\affiliation {Highly Correlated Matter Research Group, Physics Department, University of Johannesburg, P.O. Box 524, Auckland Park 2006, South Africa}
\author{A. D. Hillier}
\affiliation{ISIS Facility, Rutherford Appleton Laboratory, Chilton, Didcot, Oxon, OX11 0QX, United Kingdom}
\author{A. M. Strydom}
\affiliation {Highly Correlated Matter Research Group, Physics Department, University of Johannesburg, P.O. Box 524, Auckland Park 2006, South Africa}
\author{W. Kockelmann}
\affiliation{ISIS Facility, Rutherford Appleton Laboratory, Chilton, Didcot, Oxon, OX11 0QX, United Kingdom}
\author{B. D. Rainford}
\affiliation{Physics Department, University of Southampton, Southampton SO17 1BJ, United Kingdom}
\author{K. A. McEwen}
\affiliation{Department of Physics and Astronomy and London Centre for Nanotechnology, University College London, Gower Street, London WC1E 6BT, United Kingdom}

\date{\today}

\begin{abstract}
The results of heat capacity $C_{\rm p}(T, H)$ and electrical resistivity $\rho(T,H)$ measurements down to 0.35~K as well as muon spin relaxation and rotation ($\mu$SR) measurements on a noncentrosymmetric superconductor LaIrSi$_{3}$ are presented. Powder neutron diffraction confirmed the reported noncentrosymmetric body-centered tetragonal BaNiSn$_{3}$-type structure (space group $I4\,mm$) of LaIrSi$_{3}$. The bulk superconductivity is observed below $T_{\rm c} = 0.72(1)$~K\@. The intrinsic $\Delta C_{\rm e}/\gamma_{\rm n} T_{\rm c}  =1.09(3)$ is significantly smaller than the BCS value of 1.43, and this reduction is accounted by the $\alpha$-model of BCS superconductivity. The analysis of the superconducting state $C_{\rm e}(T)$ data by the single-band $\alpha$-model indicates a moderately anisotropic order parameter with the $s$-wave gap $\Delta(0)/k_{\rm B}T_{\rm c} = 1.54(2)$ which is lower than the BCS value of 1.764. Our estimates of various normal and superconducting state parameters indicate a weakly coupled electron-phonon driven type-I $s$-wave superconductivity in ${\rm LaIrSi_3}$. The $\mu$SR results also confirm the conventional type-I superconductivity in LaIrSi$_{3}$ with a preserved time reversal symmetry and hence a singlet pairing superconducting ground state.
\end{abstract}


\maketitle

\section{\label{Intro} Introduction}

The noncentrosymmetric superconductors (NCSs) that allow mixing between spin-singlet and spin-triplet parity and exhibit exotic superconducting properties through the antisymmetric spin-orbit coupling (ASOC) are of great interest in the current research activities on superconductivity. \cite{Bauer2012} The NCSs lack inversion symmetry in their crystal structure that leads to a non uniform lattice potential and hence introduces an antisymmetric spin-orbit coupling. The ASOC removes the spin degeneracy of conduction band electrons, i.e., the spin-up and spin-down energy bands split and the two electrons forming a Cooper pair no longer belong to the \emph{same} Fermi surface as in the conventional superconductors. An important consequence of Cooper pair formation by the electrons belonging to two \emph{different} Fermi surfaces of spin-up and spin-down bands is that the the Cooper pair wave function of NCSs can no longer be classified by its parity as a pure spin-singlet or spin-triplet pairing, instead results in a parity mixing of spin singlet-triplet states. \cite{Edelstein, Gorkov, Samokhin, Frigeri,Fujimoto2007} For centrosymmetric superconductors to which most of the known superconductors belong the spin-up and spin-down energy bands of the conduction electrons are degenerate when time reversal symmetry is conserved. The structural inversion symmetry thus has a key role in determining the superconducting properties and a number of unusual phenomena can be observed in such noncentrosymmetric materials. \cite{Edelstein, Gorkov, Samokhin, Frigeri,Fujimoto2007}

First such unusual superconducting behavior was observed in heavy fermion superconductor CePt$_{3}$Si which crystallizes in a tetragonal structure (space group $P4\,mm$) that lacks a mirror symmetry along the $c$ axis, and undergoes an antiferromagnetic transition below $T_{\rm N} =2.2$~K and becomes superconducting at the critical temperature $T_{\rm c}=0.75$~K that coexists with antiferromagnetic ordering. \cite{Bauer2004} The upper critical field $H_{\rm c2} \approx 5$~T is very high compared to the Pauli paramagnetic limiting field of $\sim 1$~T indicating a spin-triplet pairing. \cite{Bauer2004} For a spin-singlet paring Pauli paramagnetic limiting is expected. On the other hand for a system without inversion symmetry spin-triplet pairing is not permitted. \cite{Bauer2012} These contradicting situation is accounted by mixed spin singlet-triplet states of order parameter. \cite{Fujimoto2007} The irreducible representation point group for the tetragonal structure of CePt$_{3}$Si is $C_{4v}$ in which Rashba-type ASOC exists that provides the key to understand the intriguing superconducting behavior of CePt$_{3}$Si. \cite{Bauer2004,Bauer2005,Bauer2007} Following CePt$_{3}$Si many NCSs have been identified that present interesting superconducting properties including Li$_{2}$(Pd,Pt)$_{3}$B,  \cite{Togano2004,Yuan2006} CeRhSi$_{3}$,  \cite{Kimura2005,Kimura2007} CeIrSi$_{3}$, \cite{Sugitani2006} CeCoGe$_{3}$, \cite{Settai2007,Knebel2009} CeIrGe$_{3}$, \cite{Honda2010} LaNiC$_{2}$, \cite{Pecharsky1998,Hillier2009,Bonalde2011} BaPtSi$_{3}$, \cite{Bauer2009} (Rh,Ir)Ga$_{9}$,  \cite{Shibayama2007,Wakui2009} Mg$_{10}$Ir$_{19}$B$_{16}$, \cite{Klimczuk2007} Mo$_{3}$Al$_{2}$C, \cite{Karki2010} LaRhSi$_3$,\cite{Anand2011a}  Ca(Ir,Pt)Si$_{3}$, \cite{Eguchi2011} Re$_3$W, \cite{Biswas2011} Nb$_{0.18}$Re$_{0.12}$, \cite{Karki2011}  Re$_{6}$Zr, \cite{Singh2014} La(Pd,Pt)Si$_{3}$, \cite{Smidman2014}, Ca$_3$Ir$_4$Ge$_4$ \cite{vonRohr2014} etc.

The Ce-based NCSs CeRhSi$_{3}$, CeIrSi$_{3}$, CeCoGe$_{3}$ and  CeIrGe$_{3}$ crystallize with BaNiSn$_{3}$-type tetragonal structure (space group $I4\,mm$) which lacks a mirror plane symmetry along the $c$ axis and belongs to the same point group $C_{4v}$ as CePt$_{3}$Si. Thus like CePt$_{3}$Si a Rashba-type ASOC is present in these Ce$MX_3$ compounds too, leading to exotic superconducting ground state in them. \cite{Kimura2005,Kimura2007,Sugitani2006, Knebel2009, Settai2007, Honda2010, Okuda2007, Tada2010, Thamizhavel2005} Like CePt$_{3}$Si they also exhibit heavy fermion behavior and undergo a long-range antiferromagnetic ordering, however, become superconducting only under the application of pressure. \cite{Kimura2005,Kimura2007,Sugitani2006, Knebel2009, Settai2007, Honda2010, Okuda2007, Tada2010, Thamizhavel2005} The above mentioned Ce-based NCSs are situated close to a magnetic quantum critical point making it difficult to explore the effects of ASOC and inversion symmetry breaking on superconductivity. Therefore nonmagnetic Rashba-type NCSs are essential for understanding the effect and extent of ASOC on the superconducting properties of these Ce-based NCSs. The reported nonmagnetic $AMX_3$ NCSs with BaNiSn$_{3}$-type tetragonal structure include BaPtSi$_{3}$ ($T_{\rm c} = 2.25$~K), LaRhSi$_3$ [$T_{\rm c} = 2.16(8)$~K], CaIrSi$_{3}$ ($T_{\rm c} = 3.6$~K), CaPtSi$_{3}$ ($T_{\rm c} = 2.3$~K), LaPdSi$_{3}$ [$T_{\rm c} = 2.65(5)$~K] and LaPtSi$_{3}$ [$T_{\rm c} = 1.52(6)$~K].  \cite{Bauer2009,Anand2011a,Eguchi2011,Smidman2014}  All these nonmagnetic NCSs behave like conventional $s$-wave superconductor without any noticeable effect of absence of inversion symmetry in their crystal structure. Neverthless, being isostructural they  provide a direct comparison with Ce$MX_3$ NCSs and the investigations of these nonmagnetic NCSs connote the role of $4f$ moments in Ce NCSs. One important difference between the Ce$MX_3$ NCSs and these nonmagnetic NCSs is that Ce$MX_3$ exhibit superconductivity only at high pressures whereas these nonmagnetic NCSs superconduct at ambient pressure. This difference may have its origin in magnetic pairing in Ce NCSs in contrast to phonon mediated superconductivity in these nonmagnetic NCSs.

Theoretically the Rashba-type ASOC has been studied extensively and is favored for the investigations of NCSs, therefore compounds with tetragonal BaNiSn$_{3}$-type structure represents an important class of noncentrosymmetric materials. Continuing our work on BaNiSn$_{3}$-type structured materials we have performed a comprehensive study of superconducting and normal state properties of NCS LaIrSi$_3$ using heat capacity $C_{\rm p}$ and electrical resistivity $\rho$ versus temperature $T$ measurements down to 0.35~K, and muon spin relaxation and rotation ($\mu$SR) measurements down to 50~mK. The reported noncentrosymmetric body-centered tetragonal BaNiSn$_{3}$-type structure of LaIrSi$_{3}$ is confirmed by our room temperature powder XRD and neutron diffraction. Superconductivity in LaIrSi$_3$ was reported about 30 years ago with $T_{\rm c}$ between 1.9--2.7~K based on resistivity measurement. \cite{Lejay1984, Haen1985} In a recent study Okuda et al.\ reported a superconducting transition at $T_{\rm c} = 0.77$~K from the heat capacity measurement on LaIrSi$_{3}$. \cite{Okuda2007} Okuda et al.\ also carried out de Haas-van Alphen (dHvA) effect study and found that as a result of Rashba-type ASOC the Fermi surface of LaIrSi$_{3}$ splits into two Fermi surfaces (spin-up and spin-down energy bands) which are separated by about 1000 K\@.\cite{Okuda2007} In our recent investigations of superconducting properties of NCS LaRhSi$_3$ we found a conventional type-I superconductivity with preserved time reversal symmetry, however, with an unusual exponential evolution of Sommerfeld coefficient $\gamma$ with magnetic field which could be due to the reinforcement of ASOC with magnetic field.\cite{Anand2011a} Therefore in view of unusual behavior of LaRhSi$_3$ and strong effect of ASOC in LaIrSi$_3$ revealed by de Haas-van Alphen (dHvA) effect study, it was felt necessary to investigate the superconducting properties of LaIrSi$_3$ in detail, which we present in this paper.

Our $C_{\rm p}(T)$ data confirm bulk superconductivity in LaIrSi$_{3}$ below $T_{\rm c} = 0.72(1)$~K\@ in agreement with the report by Okuda et al. \cite{Okuda2007} However, the $\rho (T)$ exhibits superconductivity at a higher $T_{\rm c} = 1.45$~K apparently due to filamentary nonbulk superconductivity. The normal-state $\rho$ is metallic and well described by the Bloch-Gr\"{u}neisen model of resistivity for $T\geq1.6$~K\@. The low-$T$ normal-state $C_{\rm p}(T)$ gives electronic coefficient $\gamma_{\rm n} = 4.60(2) $~mJ/mol\,K$^{2}$ and density of states at Fermi energy ${\cal D}(E_{\rm F}) =1.95(1)$~states/eV\,f.u.\ for both spin directions, where f.u.\ stands for formula unit. A sharp jump is observed at $T_{\rm c} = 0.72(1)$~K in electronic heat capacity $C_{\rm e}(T)$ with $\Delta C_{\rm e}/\gamma_{\rm n} T_{\rm c}  =1.09(3)$ which is smaller than the BCS expected value of 1.43. Within the single-band picture the reduced value of $\Delta C_{\rm e}/\gamma_{\rm n} T_{\rm c}$ can be attributed to anisotropic energy gap in ${\rm LaIrSi_3}$. We have analyzed the superconducting state electronic heat capacity data by $\alpha$-model of BCS superconductivity \cite{Bardeen1957, Padamsee1973, Johnston2013} which suggests that the $s$-wave order parameter of ${\rm LaIrSi_3}$ is anisotropic in momentum space with an energy gap $\Delta(0)/k_{\rm B}T_{\rm c} = 1.54(2)$. We have estimated various normal and superconducting state parameters that indicate a weak-coupling type-I BCS superconductivity in dirty limit. Our $\mu$SR investigations also confirm type-I superconductivity in LaIrSi$_{3}$. Further, $\mu$SR results also show that the time-reversal symmetry is preserved as is expected for a conventional s-wave singlet pairing superconductivity. No evidence of parity mixing is observed as one would have expected in view of splitting of spin-up and spin-down energy bands revealed by dHvA study. \cite{Okuda2007}

\section{\label{ExpDetails} Experimental Details}

A polycrystalline sample of LaIrSi$_{3}$ was prepared by the standard arc melting of stoichiometric mixture of high purity elements (La:  99.9\%, Ir: 99.99\%, Si 99.999\%) on a water cooled copper hearth under the titanium gettered inert argon atmosphere with several flips to ensure homogeneity. The arc melted sample was further heat treated at 900 $^{\circ}$C for a week under the dynamic vacuum. The crystal structure was determined by the powder x-ray diffraction (XRD) using Cu K$_{\alpha}$ radiation. The heat capacity measurements were performed by the relaxation method with a physical properties measurement system (PPMS, Quantum Design, Inc.). The electrical resistivity measurements were performed by the standard four probe ac technique using the PPMS. Temperatures down to 0.35~K were attained by a $^3$He attachment to PPMS.

Powder neutron diffraction (ND) measurement was performed at room temperature using the ROTAX diffractometer at the ISIS facility of the Rutherford Appleton Laboratory, Didcot, U.K. The $\mu$SR measurements were carried out using the MuSR spectrometer at the ISIS facility with the detectors in both longitudinal and transverse configurations. A high purity silver plate was used to mount the sample which gives only a nonrelaxing muon signal. The powdered sample was mounted on silver plate using diluted GE varnish that was covered with kapton film. Temperatures down to 50~mK were achieved by cooling the sample in a dilution refrigerator. Correction coils were used to counter-effect the stray fields at the sample position to within 1 $\mu$T.

\section{\label {Crystallography} Crystallography}

\begin{figure}
\includegraphics[width=3in, keepaspectratio]{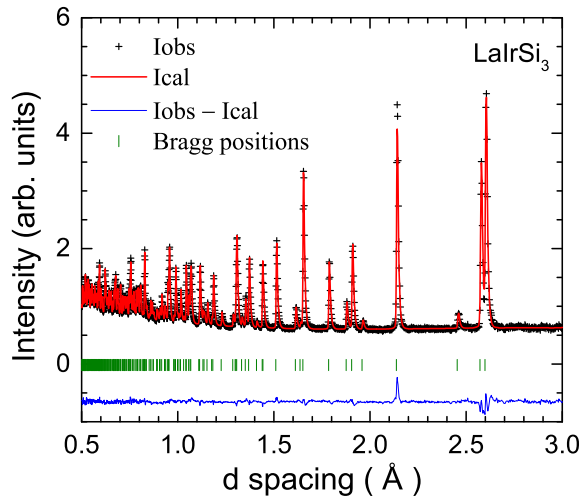}
\caption {(Color online) Powder neutron diffraction pattern of LaIrSi$_{3}$ recorded at room temperature. The solid line through the experimental points is the Rietveld refinement profile calculated for noncentrosymmetric body-centered tetragonal BaNiSn$_{3}$-type structure (space group $I4\,mm$). The short vertical bars indicate the Bragg peak positions. The lowermost curve represents the difference between the experimental and calculated intensities.}
\label{fig:ND}
\end{figure}

\begin{table}
\caption{\label{tab:XRD} Crystallographic parameters obtained from the structural Rietveld refinement of room temperature powder neutron diffraction data of LaIrSi$_{3}$. Profile reliability factor $R_{p} = 2.60\% $ and weighted profile $R$-factor $R_{wp}=2.33\%$.}
\begin{ruledtabular}
\begin{tabular}{cccccc}

\multicolumn{2}{l}{Structure} &\multicolumn{3}{l} {BaNiSn$_{3}$-type tetragonal} \\
\multicolumn{2}{l}{Space group} & \multicolumn{2}{l} {$I4\,mm$ (No. 107)}\\
\multicolumn{2}{l}{\underline{Lattice  parameters}} \\
\multicolumn{2}{l}{\hspace{0.8cm} $a$ ({\AA})}            			&  4.2784(3)  \\	
\multicolumn{2}{l}{\hspace{0.8cm} $c$ ({\AA})}          			&  9.8308(7)  \\
\multicolumn{2}{l}{\hspace{0.8cm} $V_{\rm cell}$  ({\AA}$^{3}$)} 	&  179.95(4)  \\
\\
\multicolumn{2}{l}{\underline{Atomic coordinates}} \\
Atom & Wyckoff & \hspace{-0.5 cm} $x$ & \hspace{-0.7 cm} $y$ & \hspace{-0.5 cm} $z$  & $U_{iso}$ ({\AA}$^{2}$)\\
&  position \\
La   &  2a & \hspace{-0.5 cm} 0 & \hspace{-0.7 cm} 0 & \hspace{-0.5 cm} 0 & 0.0008(3)\\
Ir   &  2a & \hspace{-0.5 cm} 0 &  \hspace{-0.7 cm} 0 & \hspace{-0.5 cm} 0.6554(2) & 0.0021(3)\\
Si1  &  2a & \hspace{-0.5 cm} 0 & \hspace{-0.7 cm} 0 & \hspace{-0.5 cm} 0.4140(3) & 0.0003(3)\\
Si2  &  4b & \hspace{-0.5 cm} 0 & \hspace{-0.7 cm} 1/2 & \hspace{-0.5 cm} 0.2624(2) & 0.0033(3)\\
\end{tabular}
\end{ruledtabular}
\end{table}

The room temperature powder XRD data were analyzed by structural Rietveld refinement using the program Fullprof. \cite{Rodriguez1993} The refinement confirmed the reported BaNiSn$_{3}$-type tetragonal structure (space group $I4\,mm$) of LaIrSi$_{3}$ and revealed the single phase nature of sample without any trace of impurity phase. The single phase nature of whole bulk of sample is further inferred from the Rietveld refinement of room temperature powder neutron diffraction data that was performed using the program GSAS.\cite{Larson2004} Neutron diffraction pattern and refinement profile for noncentrosymmetric body-centered tetragonal BaNiSn$_{3}$-type structure are shown in Fig.~\ref{fig:ND}. While refining no improvement in the fit quality was observed upon refining the occupancies of atomic positions, and within the error bar the atomic occupancies were found to be unity, therefore in the final refinement we fixed the occupancies to unity.  The crystallographic parameters obtained from the refinement of powder neutron diffraction are listed in Table~\ref{tab:XRD}. Both ND and XRD data gave similar crystallographic parameters and agree well with the literature values. \cite{Lejay1984,Engel1983}

\begin{figure}
\includegraphics[width=3in, keepaspectratio]{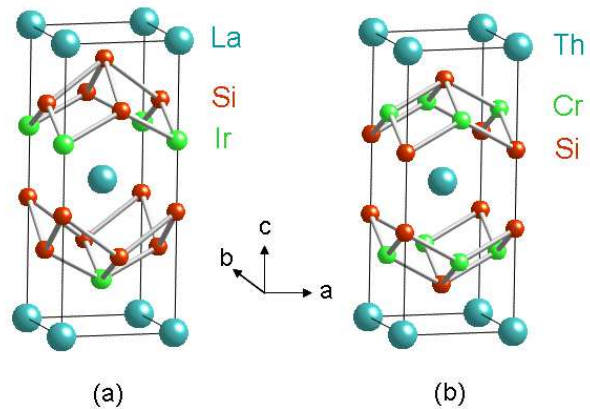}
\caption{(Color online) Comparison of (a) BaNiSn$_{3}$-type noncentrosymmetric body-centered tetragonal structure ($I4\,mm$) of LaIrSi$_{3}$ and (b) ${\rm ThCr_2Si_2}$-type body-centered tetragonal crystal structure ($I4/mmm$).}
\label{fig:structure}
\end{figure}

The BaNiSn$_{3}$-type body-centered tetragonal structure (space group $I4\,mm$) of LaIrSi$_{3}$ is illustrated in Fig.~\ref{fig:structure} and is compared with the common ThCr$_{2}$Si$_{2}$-type body-centered tetragonal structure (space group $I4/mmm$). Like ThCr$_{2}$Si$_{2}$-type structure the BaNiSn$_{3}$-type structure is also a layered structure and a ternary derivative of BaAl$_{4}$-type structure. \cite{Parthe1983} The $R$ (La, Th) atoms occupy identical positions in both structures and form body-centered tetragonal sublatttice. However, they differ in the positions of $T$ (Ir, Cr) and Si atoms. The $T$ atoms form square sublattice in $ab$ plane in both the structures but they are rotated by $45^\circ$ in $ab$ plane with respect to each other. Further, in ThCr$_{2}$Si$_{2}$-type structure all the Si atoms occupy a single crystallographic site whereas in BaNiSn$_{3}$-type structure the Si atoms occupy two different sites and hence the stacking order of $T$ and Si layers along the $c$ axis is different in the two structures. The structural difference in the two structures is evident from Fig.~\ref{fig:structure}. It is seen that the BaNiSn$_{3}$-type structure is not symmetric about the $R$ plane and there is a loss of mirror plane along the $c$ axis in BaNiSn$_{3}$-type structure which is present in the ThCr$_{2}$Si$_{2}$-type structure. The ThCr$_{2}$Si$_{2}$-type structure is centrosymmetric whereas the BaNiSn$_{3}$-type structure is noncentrosymmetric.

\section{\label{Sec:LaIrSi3_Rho} Electrical Resistivity}

\begin{figure}
\includegraphics[width=3in]{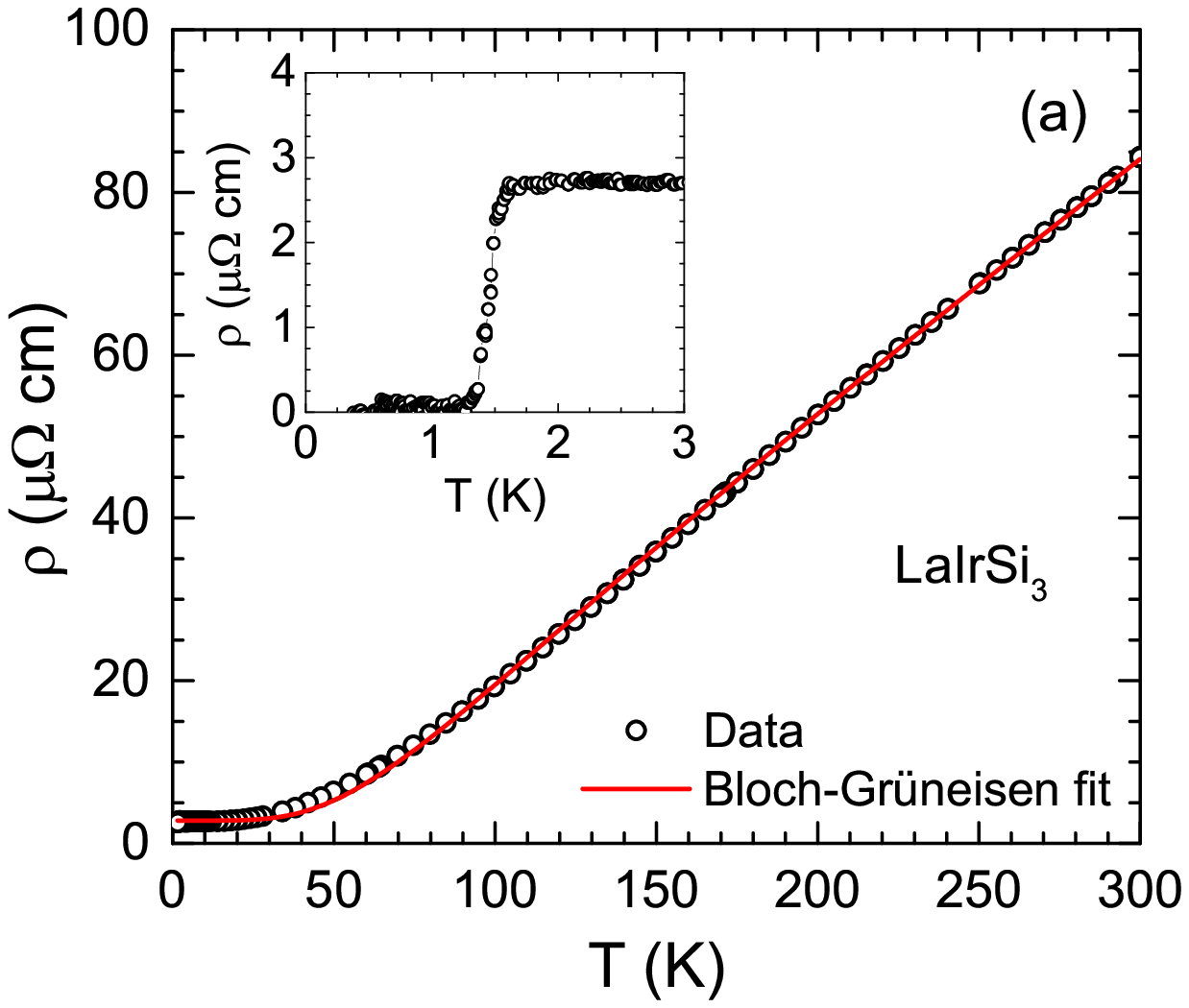} \vspace{0.1in}
\includegraphics[width=3in]{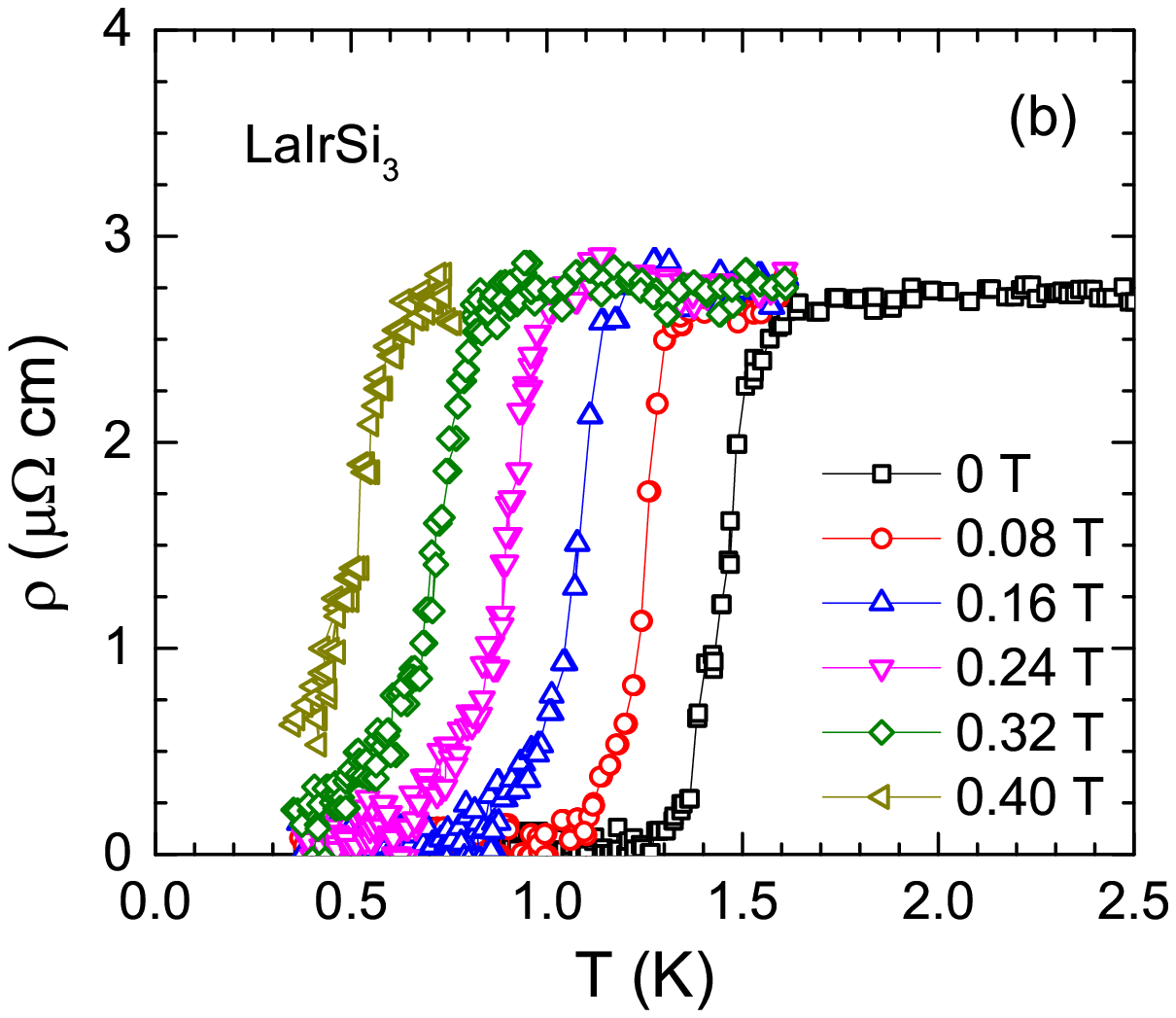}
\caption{(Color online) (a) Electrical resistivity $\rho$ of ${\rm LaIrSi_3}$ as a function of temperature~$T$ for 1.6~K~$\leq T \leq$~300~K measured in applied magnetic field $H=0$. The red solid curve is a fit of $\rho(T)$ data by the Bloch-Gr\"{u}neisen model. Inset: Expanded view of $\rho(T)$ below 3~K to show the superconductivity. (b) The $\rho(T)$ of ${\rm LaIrSi_3}$ for 0.35~K~$\leq T \leq$~2.5~K showing the superconducting transitions for different values of $H$.}
\label{fig:LaIrSi3_rho}
\end{figure}

The electrical resistivity $\rho$ of ${\rm LaIrSi_3}$ as a function of $T$ for 0.35~K~$\leq T \leq$~300~K measured at applied magnetic field $H = 0$ is shown in Fig.~\ref{fig:LaIrSi3_rho}(a). A metallic behavior is inferred from the $T$ dependence of $\rho$, the $\rho$ decreases with decreasing $T$, becomes nearly constant in the low-$T$ limit below 25~K and undergoes a sharp transition to a zero resistance state due to the occurrence of superconductivity. It is seen that onset of superconductivity takes place at $T_{\rm c\,onset} \approx 1.6$~K and zero resistivity state is reached at $T_{\rm c\,0} \approx 1.3$~K\@. Thus a $T_{\rm c} = 1.45$~K (defined as the mid point of the transition) is obtained from the resistivity data. The $\rho(T)$ data at various $H$ for $0 \leq H \leq 0.4$~T are shown in Fig.~\ref{fig:LaIrSi3_rho}(b). It is seen that the $T_{\rm c}$ decreases with increasing $H$, at $H = 0.32$~T, $T_{\rm c}$ reduces to 0.70~K from 1.45~K at $H=0$.

From Fig.~\ref{fig:LaIrSi3_rho}(a) the residual resistivity before entering the superconducting state is $\rho_0 = 2.7~\mu \Omega$\,cm and a residual resistivity ratio  \mbox{RRR~$\equiv \rho(300\,{\rm K}) / \rho(1.6\,{\rm K}) \approx 31$} . The low value of $\rho_0$ and high value of RRR indicate good sample quality. In the normal state, the $\rho(T\geq1.6~{\rm K})$ data are well described by the Bloch-Gr\"{u}neisen (BG) model of resistivity due to the scattering of conduction electrons by longitudinal acoustic lattice vibrations. \cite{Blatt1968} We fitted our normal-state $\rho(T)$ data by
\begin{equation}
\rho(T) = \rho_0 + \rho_{\rm BG},
\label{eq:BG_fit}
\end{equation}
where
\begin{equation}
\rho_{\rm BG}(T/\Theta_{\rm R})= 4 \mathcal{R} \left( \frac{T}{\Theta _{\rm{R}}}\right)^5 \int_0^{\Theta_{\rm{R}}/T}{\frac{x^5}{(e^x-1)(1-e^{-x})}dx},
 \label{eq:BG}
\end{equation}
represents the BG resistivity. The $\Theta_{\rm R}$ is the Debye temperature from the resistivity data and $\mathcal{R}$ is a material-dependent prefactor.

Our fit of $\rho(T)$ data in 1.6~K~$\leq T \leq$~300~K by BG model is shown by the solid red curve in Fig.~\ref{fig:LaIrSi3_rho}(a) where we used an analytic Pad\'e approximant fitting function for $\rho_{\rm BG}$ from Ref.~\onlinecite{Goetsch2012}. From the fitting of $\rho(T)$ data we obtain $\rho_0 = 2.81(2)~\mu \Omega$\,cm, $\Theta_{\rm{R}} = 331(2)$~K and $\mathcal{R} = 95.7~\mu \Omega$\,cm. Further details about the fitting of $\rho(T)$ by the Bloch-Gr\"{u}neisen model of resistivity can be found in Refs.~\onlinecite{Goetsch2012} and \onlinecite{Anand2012a}.

\section{\label{Sec:LaIrSi3_HC} Heat Capacity}

\begin{figure}
\includegraphics[width=3in]{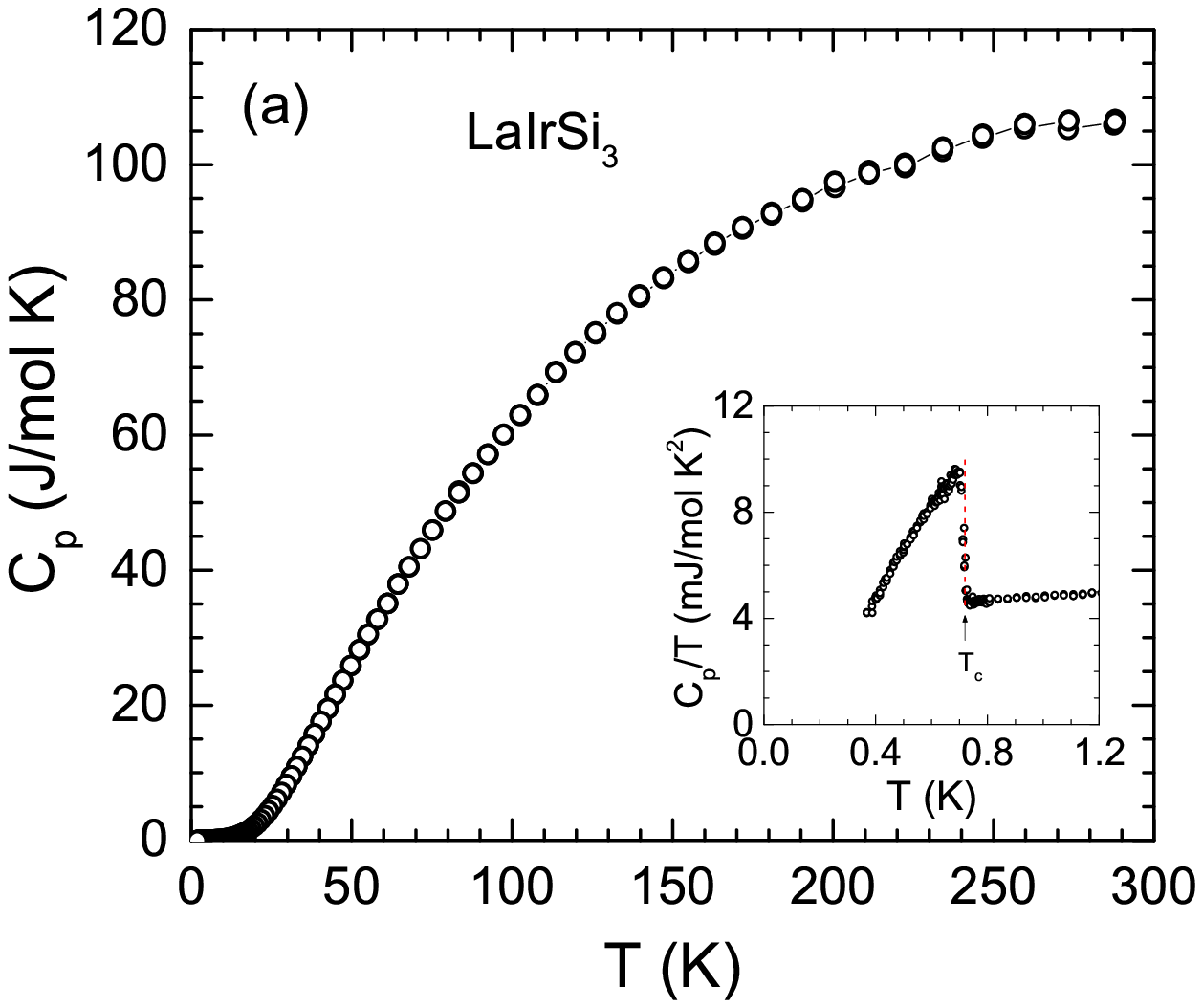}\vspace{0.1in}
\includegraphics[width=3in]{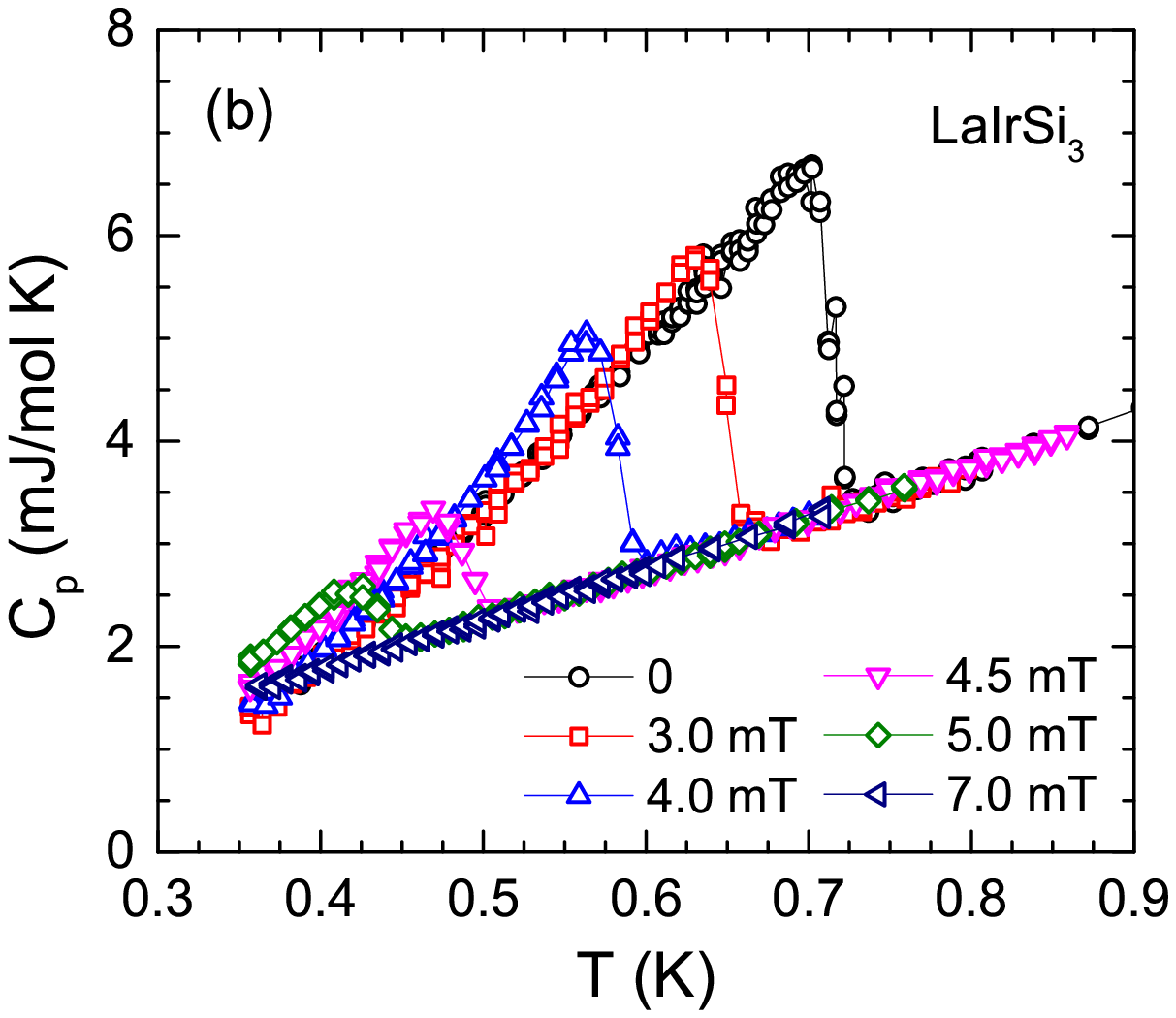}\vspace{0.1in}
\includegraphics[width=3in]{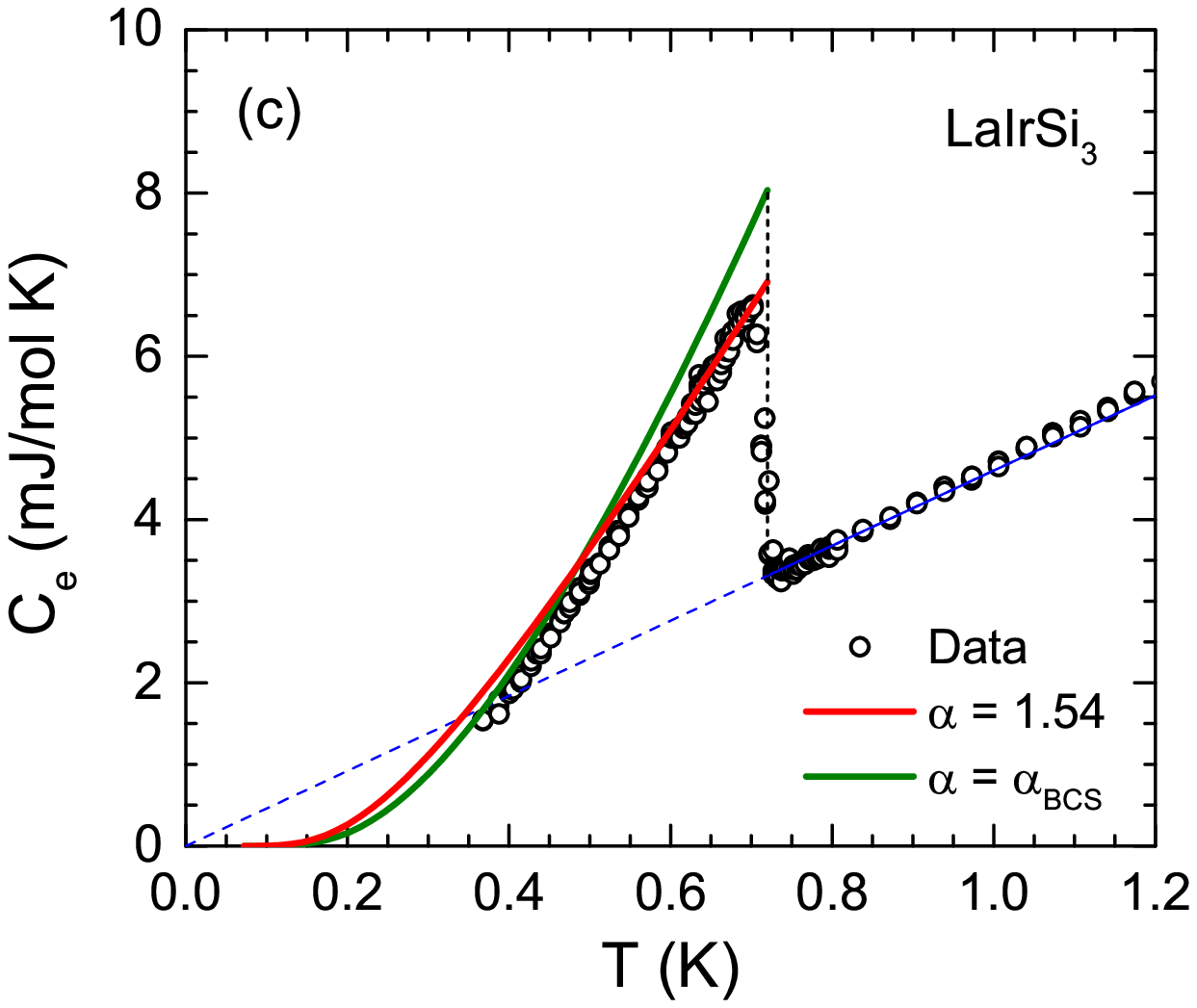}
\caption{(Color online) (a) Heat capacity $C_{\rm p}$ of ${\rm LaIrSi_3}$ as a function of temperature $T$ for 1.8~K~$\leq T \leq$~300~K measured in zero magnetic field. Inset: $C_{\rm p}/T$ vs. $T$ for $0.35~{\rm K} \leq T \leq 1.2$~K\@. The dotted red line mark the superconducting transition temperature $T_{\rm c}$. (b) $C_{\rm p}(T)$ for 0.35~K~$\leq T \leq$~0.9~K measured in different applied magnetic fields. (c) Electronic contribution $C_{\rm e}$ to zero field heat capacity as a function of temperature $T$. The solid red curve is the theoretical prediction of the $\alpha$-model for $\alpha = \Delta(0)/k_{\rm B}T_{\rm c} = 1.54$.  The BCS prediction for the $\alpha_{\rm BCS} = 1.764$ is also shown for comparison.}
\label{fig:LaIrSi3_HC}
\end{figure}

The heat capacity $C_{\rm p}$ of ${\rm LaIrSi_3}$ as a function of $T$ for 0.35~K~$\leq T \leq$~300~K measured at $H = 0$ is shown in Fig.~\ref{fig:LaIrSi3_HC}(a). As shown in the inset a sharp jump is observed in $C_{\rm p}$ due to the superconducting transition at $T_{\rm c} = 0.72(1)$~K. The observation of such a sharp jump in $C_{\rm p}(T)$ indicates the occurrence of bulk superconductivity in ${\rm LaIrSi_3}$. The $C_{\rm p}(T)$ data measured at different magnetic fields are shown in Fig.~\ref{fig:LaIrSi3_HC}(b). It is seen that the jump $\Delta C_{\rm p}$ in $C_{\rm p}$ as well as $T_{\rm c}$ decrease with the increasing $H$. The $T_{\rm c}$ is found to decrease to 0.44(2)~K at $H=5.0$~mT from its value of $T_{\rm c} = 0.72(1)$~K at $H=0$, and is suppressed to a temperature below 0.35~K by a field of $H = 7.0$~mT\@. The suppression of $T_{\rm c}$ with $H$ for $C_{\rm p}(T,H)$ is very different from that observed for the $\rho(T, H)$ data above where superconductivity survives even at an applied field of 0.4~T\@.

The low temperature normal state heat capacity data are well described by $C_{\rm p}(T)=\gamma_{\rm n} T + \beta T^3$. The normal state Sommerfeld electronic heat capacity coefficient $\gamma_{\rm n}$ is estimated by fitting the normal-state $C_{\rm p}(T)$ data measured at $H=0$ (in $0.75~{\rm K} \leq T \leq 3.8$~K) and at $H=10.0$~mT (in $0.35~{\rm K} \leq T \leq 0.9$~K) simultaneously. The simultaneous linear fit of $C_{\rm p}(T)/T$ versus $T^2$ by $C_{\rm p}(T)/T=\gamma_{\rm n} + \beta T^2$ in $0.35~{\rm K} \leq T \leq 3.8$~K yield $\gamma_{\rm n} = 4.60(2)$~mJ/mol\,K$^2$ and $\beta= 0.17(1)$~mJ/mol\,K$^4$. The coefficient $\beta$ according to the relation \cite{Kittel2005} $\Theta_{\rm D} = (12 \pi^{4} R p/5 \beta )^{1/3}$, where $R$ is the molar gas constant and $p=5$ is the number of atoms per formula unit (f.u.), gives the Debye temperature $\Theta_{\rm D}= 385(8)$~K\@. The experimental $C_{\rm p}(T=300~{\rm K}) \approx 106$~J/mol\,K does not reach the Dulong-Petit high-$T$ limit of the lattice heat capacity $C_{V}$ = $3pR$ = 15$R$ = 124.7 J/mole\,K.

The coefficient $\gamma_{\rm n}$ can be used to estimate the density of states at the Fermi level ${\cal D}(E_{\rm F})$ which according to the relation \cite{Kittel2005} $\gamma_{\rm n} = (\pi^2 k_{\rm B}^2/3)\, {\cal D}(E_{\rm F})$ gives ${\cal D}(E_{\rm F}) = 1.95(1)$~states/eV\,f.u.\ for both spin directions. This ${\cal D}(E_{\rm F})$ contains the quasi-particle mass enhancement by many-body electron-phonon interaction and is related to the bare density of states ${\cal D}_{\rm band}(E_{\rm F})$ by \cite{Grimvall1976} ${\cal D}(E_{\rm F}) = (1 + \lambda_{\rm e-ph}){\cal D}_{\rm band}(E_{\rm F})$, where $\lambda_{\rm {e-ph}}$ is the electron-phonon coupling constant that can be estimated from $\Theta_{\rm D}$ and $T_{\rm c}$ using the McMillan's relation \cite{McMillan1968}
\begin{equation}
\lambda_{\rm {e-ph}}= \frac {1.04+\mu^{\ast} \ln(\Theta_{\rm D}/1.45\,T_{\rm c})} {(1-0.62\mu^{\ast})\ln(\Theta_{\rm D}/1.45\,T_{\rm c}) - 1.04}.
\label{eq:lambda}
\end{equation}
Here $\mu^{\ast}$ is the repulsive screened Coulomb parameter usually assigned as $\mu^{\ast} = 0.13$. For ${\rm LaIrSi_3}$ we have $T_{\rm c} = 0.72$~K and $\Theta_{\rm D}= 385$~K which together with $\mu^{\ast}=0.13$, according to Eq.~(\ref{eq:lambda}) gives $\lambda_{\rm {e-ph}} = 0.41$ . The small value of $\lambda_{\rm {e-ph}}$ implies a weak-coupling superconductivity in ${\rm LaIrSi_3}$. This value of $\lambda_{\rm {e-ph}}$ combined with ${\cal D}(E_{\rm F}) = 1.95(1)$~states/eV\,f.u.\ for both spin directions gives ${\cal D}_{\rm band}(E_{\rm F})= 1.38(1)$~states/eV\,f.u.\ for both spin directions. The effective mass $m^\ast$ of the quasi-particle can be obtained from $ m^\ast = (1 + \lambda_{\rm e-ph})\,m_{\rm band}^\ast$ which gives $m^\ast  = 1.41\, m_{\rm e}$ assuming the effective band mass $m_{\rm band}^\ast = m_{\rm e}$, the free electron mass.

The density of states and hence $\gamma_{\rm n}$ can be further used to estimate the Fermi velocity $v_{\rm F}$ which is related to ${\cal D}(E_{\rm F})$ by \cite{Kittel2005} $v_{\rm F} = (\pi^2 \hbar^3/m^{\ast 2} V_{\rm f.u.}) {\cal D}(E_{\rm F})$, where $\hbar$ is Planck's constant divided by $2\pi$ and $V_{\rm f.u.} = V_{\rm cell}/2$ is the volume per formula unit. We thus estimate $v_{\rm F}  = 9.49 \times 10^7~{\rm cm/s}$ for ${\rm LaIrSi_3}$ using the above estimated ${\cal D}(E_{\rm F})$ and $m^\ast$. The $v_{\rm F}$ together with $\rho_0$ can be used to estimate the mean free path $\ell$, as $\ell = v_{\rm F}\tau$ and the mean free scattering time $\tau= m^\ast /n e^2 \rho_0$, with the conduction carrier density $n = m^{\ast 3} v_{\rm F}^3/3\pi^2\hbar^3$ assuming a spherical Fermi surface. \cite{Kittel2005} Combining all these
\be
\ell = 3\pi^2 \left(\frac{\hbar}{e^2\rho_0}\right)\left(\frac{ \hbar}{m^\ast v_{\rm F}}\right)^2.
\label{eq:lvF}
\ee
which for $\rho_0 = 2.7~\mu\Omega$\,cm and above estimated $v_{\rm F}$ and $m^\ast$ gives $\ell = 33.7~{\rm nm}$.

\section{\label{Sec:LaIrSi3_SC_Properties} Superconducting state properties}

The electronic contribution to the heat capacity $C_{\rm e}(T)$ after subtracting off the phonon contribution from the measured $C_{\rm p}(T)$ data, i.e. $C_{\rm e}(T) = C_{\rm p}(T) - \beta T^3$ is shown in Fig.~\ref{fig:LaIrSi3_HC}(c) clearly showing the sharp jump in $C_{\rm e}$ at $T_{\rm c}$. The jump $\Delta C_{\rm e}$ in $C_{\rm e}$ at $T_{\rm c}$ is found to be $\Delta C_{\rm e} = 3.6(1)$~mJ/mol\,K corresponding to the vertical dotted line at $T_{\rm c}$ in Fig.~\ref{fig:LaIrSi3_HC}(c). This gives $\Delta C_{\rm e}/ \gamma_{\rm n} T_{\rm c} = 1.09(3)$ for $\gamma_{\rm n} = 4.60(2)$~mJ/mol\,K$^2$ and $T_{\rm c}=0.72(1)$~K which is significantly smaller than the BCS value of $\Delta C_{\rm e}/ \gamma_{\rm n} T_{\rm c} =1.43$ in the weak-coupling limit.\cite{Tinkham1996} Presence of a residual heat capacity due to small impurity/nonsuperconducting phase can lead to a reduced $\Delta C_{\rm e}/ \gamma_{\rm n} T_{\rm c}$. However, in view of the fact that the jump in $C_{\rm e}(T)$ at the superconducting transition is very sharp, the entire sample seems to be superconducting without any residual  $\gamma$, which in turn suggests that the reduction in $\Delta C_{\rm e}/ \gamma_{\rm n} T_{\rm c}$ is intrinsic. In a single-band model, such a reduction can be caused by the presence of anisotropic superconducting energy gap (order parameter) in momentum space. \cite{Johnston2013} That the reduction in $\Delta C_{\rm e}/ \gamma_{\rm n} T_{\rm c}$ is intrinsic will be clear from our analysis of the superconducting state data by single-band $\alpha$-model of BCS superconductivity \cite{Bardeen1957, Padamsee1973, Johnston2013} below which is applicable to the system with $\Delta C_{\rm e}/ \gamma_{\rm n} T_{\rm c} \neq 1.43$.

In the so-called $\alpha$-model of BCS superconductivity in order to fit the superconducting state thermodynamic data, the $\alpha_{\rm BCS} \equiv \Delta(0)/k_{\rm B}T_{\rm c} = 1.764$ is replaced by a variable $\alpha$. \cite{Padamsee1973, Johnston2013}  The $\alpha$ is determined from the jump $\Delta C_{\rm e}$  at $T_{\rm c}$ according to the relation \cite{Johnston2013}
\be
\frac{\Delta C_{\rm e}(T_{\rm c})}{\gamma_{\rm n}T_{\rm c}} = 1.426\left(\frac{\alpha}{\alpha_{\rm BCS}}\right)^2
\ee
which for $\Delta C_{\rm e}/ \gamma_{\rm n} T_{\rm c} = 1.09(3)$ gives $\alpha = 1.54(2)$. This value of $\alpha$ is significantly smaller than the BCS value of~1.764. The temperature dependence of $\alpha$-model superconducting state heat capacity $C_{\rm e}(T)$ calculated for $\alpha = 1.54$ is shown by the solid red curve in Fig.~\ref{fig:LaIrSi3_HC}(c) together with that of BCS prediction for $\alpha = \alpha_{\rm BCS}$. A reasonable agreement is observed between the $\alpha$-model prediction and the superconducting state $C_{\rm e}(T)$ data which supports the applicability of $\alpha$-model and in turn indicates that the $s$-wave order parameter of ${\rm LaIrSi_3}$ is anisotropic in momentum space. The details about the fitting of $C_{\rm e}(T)$ data by $\alpha$-model of BCS superconductivity can be found in Refs.~\onlinecite{Johnston2013} and \onlinecite{Anand2013a}. The lack of perfect agreement between the $\alpha$-model prediction and the experimental data may indicate that the anisotropy of gap is not well accounted, as the simplified $\alpha$-model does not account for the energy dependence of the gap function or for a complex $\Delta$.

\begin{figure}
\includegraphics[width=3in, keepaspectratio]{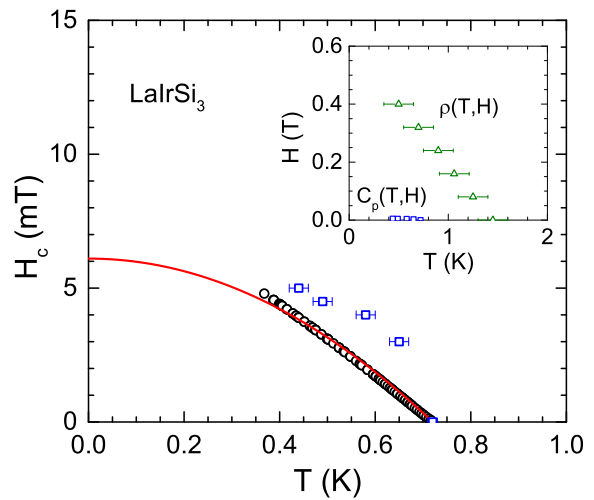}\vspace{0.1in}
\caption{\label{fig:LaIrSi3_Hcritical} (Color online) Thermodynamic critical field $H_{c}$ of LaIrSi$_{3}$ as a function of temperature $T$  obtained from the experimental electronic heat capacity $C_{\rm e}(T)$ data. The $H$ dependent $T_{\rm c}$ obtained from the heat capacity $C_{\rm p}(T, H)$ data  in Fig.~\ref{fig:LaIrSi3_HC}(b) are also shown by open squares. The solid line represents the parabolic fit to $H_{\rm c}(T) = H_{\rm c}(0)[1 - (T/T_{c})^{2}]$ as discussed in the text. Inset: $H$--$T$ phase diagram obtained from the electrical resistivity $\rho(T, H)$ in Fig.~\ref{fig:LaIrSi3_rho}(b) together with $T_{\rm c}(H)$ from $C_{\rm p}(T, H)$.}
\end{figure}

The thermodynamic critical field $H_{\rm c}$ is estimated from the zero-field $C_{\rm e}(T)$ data by integrating the entropy difference between the superconducting and normal states, \cite{Tinkham1996, DeGennes1966} $H_{\rm c}^{2}(T) = 8\pi\int_{T}^{T_{\rm c}}[S_{\rm en}(T^\prime)-S_{\rm es}(T^\prime)] dT^\prime $, where $S_{\rm en}$ and $S_{\rm es}$ are the electronic entropy of normal and superconducting states, respectively, with $S_{\rm e}(T^{\prime}) = \int_0^{T^\prime}[C_{\rm e}(T^{\prime\prime})/T^{\prime\prime})]dT^{\prime\prime}$. The $H_{\rm c}(T)$ obtained from the zero-field $C_{\rm e}(T)$ this way is shown in Fig.~\ref{fig:LaIrSi3_Hcritical}. The $T$ dependence of $H_{\rm c}$ can be approximated to the conventional relation $H_{\rm c}(T) = H_{\rm c}(0)[1 - (T/T_{c})^{2}]$, which gives $H_{\rm c}(0) =6.10(3)$~mT. The solid red curve in Fig.~\ref{fig:LaIrSi3_Hcritical} represents the fit of $H_{\rm c}(T)$ data with this expression.

The experimental $H_{\rm c}(0)$ obtained so is somewhat higher than the theoretical $H_{\rm c}(0)$ which for $\alpha$-model is given by \cite{Johnston2013}
\be
\frac{H_{\rm c}(0)}{\left(\gamma_{\rm nV}T_{\rm c}^2\right)^{1/2}} = \sqrt{\frac{6}{\pi}}\ \alpha \approx 1.382\,\alpha,
\label{Eq:HcFromGammaTc}
\ee
where the Sommerfeld coefficient per unit volume $\gamma_{\rm nV}$ is in units of ${\rm erg/cm^3\,K^2}$. From this relation for $\alpha = 1.54$ we obtain $H_{\rm c}(0) = 4.5$~mT which is little lower that the $H_{\rm c}(0) =6.10(3)$~mT obtained above. Even for $\alpha_{\rm BCS} = 1.764$, Eq.~(\ref{Eq:HcFromGammaTc}) gives a lower $H_{\rm c}(0)= 5.1$~mT. The reason for this discrepancy between the experimental and theoretical values of $H_{\rm c}(0)$ is not clear. We suspect that this might be the result of a nonspherical/anisotropic Fermi surface in LaIrSi$_{3}$ which is not properly accounted by Eq.~(\ref{Eq:HcFromGammaTc}).

Further, as can be seen from  Fig.~\ref{fig:LaIrSi3_Hcritical}, the $H_{\rm c}(T)$ obtained from the zero-field $C_{\rm e}(T)$ data and $H$--$T$ phase diagram determined from the $H$ dependent $T_{\rm c}$ from $C_{\rm p}(T, H)$ data in Fig.~\ref{fig:LaIrSi3_HC}(b) both give very low critical fields. For a type-II superconductivity the $T_{\rm c}(H)$ obtained from $C_{\rm p}(T, H)$ gives the upper critical field $H_{\rm c2}$ that is usually much higher than the thermodynamic critical field, which is not the present case. The fact that $H_{\rm c2}(T)$ is close to $H_{\rm c}(T)$ may suggest a type-I superconductivity in LaIrSi$_{3}$ or a type 1.5 behavior that may arise from the presence of split spin-up and spin-down energy bands similar to what has been observed in two-band superconductor MgB$_2$. \cite{Moshchalkov2009}

The $H$ dependent $T_{\rm c}$ from $\rho(T, H)$ data in Fig.~\ref{fig:LaIrSi3_rho}(b) is shown in the inset of Fig.~\ref{fig:LaIrSi3_Hcritical} which show very different behavior than the $T_{\rm c}(H)$ from $C_{\rm p}(T, H)$ data. This is consistent with the observation of different $T_{\rm c}$'s in resistivity and heat capacity measurements in zero field mentioned above, apparently due to the filamentary/surface superconductivity that sets in at a temperature higher than the bulk superconductivity. Our estimate of Ginzburg-Landau parameter $\kappa $ below gives $\kappa = 0.55$ which suggests a type-I behavior. Usually for a type-I superconductor one do not expect a filamentary or surface superconductivity. However, surface superconductivity is predicted for a type-I superconductor with $\kappa $ values between $1/\surd 2$ and $1/2.39$. \cite{Saint-James1963,Strongin1964} Indeed our estimated $\kappa $ lies between these limits and we can expect a surface superconductivity in LaIrSi$_{3}$. The critical field associated with the surface superconductivity is given by \cite{Saint-James1963,Strongin1964}   $H_{\rm c3} = 2.39\, \kappa H_{\rm c}$, accordingly for LaIrSi$_{3}$ we estimate $H_{\rm c3} = 8.0$~mT which is much smaller than the observed upper critical field from resistivity measurement (inset of Fig.~\ref{fig:LaIrSi3_Hcritical}). Thus the observed $T_{\rm c}(H)$ from $\rho(T, H)$ could not be understood to arise from surface superconductivity. The reason for such high critical field in resistivity measurement of LaIrSi$_{3}$ is not clear. The type I superconductor LaPdSi$_3$ was also found to exhibit a higher critical field in resistivity measurement compared to that in heat capacity.\cite{Smidman2014} In a recent study N. Kimura et al. \cite{Kimura2014}  found a similar high critical field from the resistivity measurement on type-I superconductor LaRhSi$_{3}$ where they argue that this behavior may be a common feature in noncentrosymmetric superconductors. However, we are not aware of any theoretical discussion of this aspect of noncentrosymmetric superconductors, and this hypothesis needs to be tested theoretically.

\begin{table}
\caption{\label{tab:SCParams} Measured and derived superconducting and relevant normal state parameters for the noncentrosymmetric superconductor ${\rm LaIrSi_3}$.}
\begin{ruledtabular}
\begin{tabular}{lc}
$T_{\rm c}$ (K)                                         &  0.72(1)   \\
$\gamma_{\rm n}$ (mJ/mol\,K$^{2}$)                      & 4.60(2) \\
$\Theta_{\rm D}$ (K)                                    & 385(8) \\
$\lambda_{\rm e-ph}$                                    & 0.41   \\
$\Delta C_{\rm e}$ (mJ/mol\,K)                          &  3.6(1) \\
$\Delta C_{\rm e}/\gamma_{\rm n} T_{\rm c}$             &  1.09(3) \\
$\alpha \equiv \Delta(0)/k_{\rm B}T_{\rm c}$ (from $\Delta C_{\rm e}/\gamma_{\rm n} T_{\rm c}$) &   1.54(2)\\
$\Delta(0)/k_{\rm B}$~(K) 				                & 1.11	\\
$H_{\rm c}(T=0)$ (mT)                                   &  6.10(3) \\
$\kappa_{\rm GL}$                                       &  0.55 \\
$\xi(0)$ (nm)                                         &  373 \\
$\xi_0$ (nm)                                            &  2081 \\
$\ell~$ (nm)                                            &  33.7  \\
$\lambda_{\rm L}(0)$~(clean limit) (nm)      &  25.9   \\
$\lambda_{\rm eff}(0)$~(dirty limit) (nm)    &  205 \\
\end{tabular}
\end{ruledtabular}
\end{table}

Our estimate of the superconducting London penetration depth in the clean limit at $T=0$, $\lambda_{\rm L}(0)$ from $v_{\rm F}$ using the relation \cite{Tinkham1996}
\be
\lambda_{\rm L}(0)^{2} = \frac{m^{\ast}c^2}{4 \pi n e^2} = \frac{3\pi c^2 \hbar^3}{4 m^{\ast 2} e^2 v_{\rm F}^3} ,
\label{Eq:lambda0}
\ee
where $c$ is the speed of light in vacuum, gives $\lambda_{\rm L}(0) = 25.9~{\rm nm}$ for $v_{\rm F}  = 9.49 \times 10^7~{\rm cm/s}$. The BCS coherence length $\xi_0$ can be obtained from $v_{\rm F}$ and energy gap $ \Delta(0)$ which for the $\alpha$-model is given by \cite{Johnston2013}
\be
\xi_0 = \frac{\hbar v_{\rm F}}{\pi \Delta(0)} = \left(\frac{1}{\pi \alpha}\right)\frac{\hbar v_{\rm F}}{k_{\rm B} T_{\rm c}}.
\label{eq:xivF}
\ee
This gives $\xi_0 = 2081$~nm for $\alpha=1.54$ and $v_{\rm F}  = 9.49 \times 10^7~{\rm cm/s}$. We see that $\xi_0$ is much larger than the above estimated mean free path $\ell=33.7$~nm,  $\ell/\xi_{0} \approx 0.016 \ll 1$, suggesting that the superconductivity in LaIrSi$_{3}$ is in the dirty-limit. In the dirty limit, the Ginzburg-Landau parameter $\kappa_{\rm GL} = 0.715\,\lambda_{\rm L}(0)/\ell$,\cite{Tinkham1996}  which gives $\kappa_{\rm GL} = 0.55< 1/\surd2$, as expected for a type-I superconductivity. This value of $\kappa_{\rm GL}$ is close to the value obtained using the dirty-limit relation for a fully gapped (isotropic) superconductor, \cite{Orlando1979} $\kappa_{\rm GL} = 7.49 \times 10^{3} \rho_{0} \surd \gamma_{\rm nV}$, with $\rho_{0}$ in $\Omega$~cm, which gives $\kappa_{\rm GL} = 0.59$.  Both estimates of $\kappa_{\rm GL}$ consistently indicate a type-I superconductivity in LaIrSi$_{3}$.

The effective magnetic penetration depth $\lambda_{\rm eff}$ can be estimated using the dirty limit relation \cite{Tinkham1996}
\be
\lambda_{\rm eff}(0) = \lambda_{\rm L}(0)\sqrt{1+\frac{\xi_0}{\ell}}\qquad {\rm (dirty\ limit)}.
\label{eq:lambda_eff2}
\ee
which gives $\lambda_{\rm eff}(0) = 205$~nm. Then using the relation $\kappa_{\rm GL} = \lambda_{\rm eff}(0)/\xi(0)$ we estimate the Ginzburg-Landau coherence length $\xi(0)$ which for $\kappa_{\rm GL} = 0.55$ yields $\xi(0)= 373$~nm. The measured and derived superconducting parameters of ${\rm LaIrSi_3}$ are listed in Table~\ref{tab:SCParams}.

\section{\label{Sec:muSR} Muon spin relaxation and rotation}

\begin{figure}
\includegraphics[width=3in, keepaspectratio]{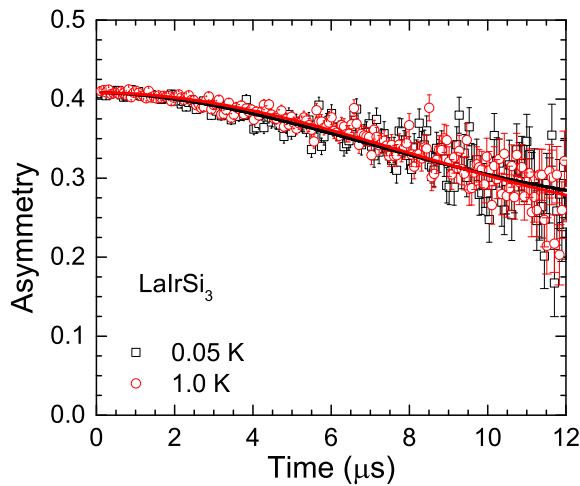}
\caption{\label{fig:musr1} (Color online) Zero field $\mu$SR spectra of ${\rm LaIrSi_3}$ measured in longitudinal geometry at temperatures above (1.0~K) and below (0.05~K) the superconducting $T_{\rm c}$. The solid curves are the fits by Gaussian Kubo-Toyabe function in Eq.~(\ref{eq:KT}).}
\end{figure}

The time evolution of muon spin relaxation in zero field (ZF) is shown in Fig.~\ref{fig:musr1} for both at $T$ above and below the bulk $T_{\rm c}$. It is evident from the ZF $\mu$SR spectra that there is no noticeable change in the relaxation rates at 1.0~K ($>T_{\rm c}$) and 0.05~K ($<T_{\rm c}$). This indicates that the time reversal symmetry remains preserved upon entering the superconducting state. The ZF $\mu$SR spectra are best described by the Gaussian Kubo-Toyabe function,
\begin{equation}
\label{eq:KT}
 G_z(t)=A_0\left[\frac{1}{3}+\frac{2}{3}\left(1-\sigma^2 t^2 \right){\rm e}^{ -\sigma ^2 t^2/2}\right] {\rm e}^{-\lambda t} + A_{\rm BG},
\end{equation}
where $A_0$ is the initial asymmetry, $\sigma$ and $\lambda$ are the depolarization rates, and $A_{\rm BG}$ is the time-independent background contribution. $\sigma$ accounts for the Gaussian distribution of static fields from nuclear moments [the local field distribution width $ \langle H_{\mu} \rangle = \sigma / \gamma_{\mu}$ with muon gyromagnetic ratio $\gamma_{\mu}$ = 135.53 MHz/T] and $\lambda$ accounts for the electronic moments.  The fits of $\mu$SR spectra in Fig.~\ref{fig:musr1} by the decay function in Eq.~(\ref{eq:KT}) give $\sigma = 0.074(1)~\mu $s$^{-1}$ and $\lambda = 0.009(3)~\mu$s$^{-1}$ at 1.0~K and $\sigma = 0.074(1)~\mu $s$^{-1}$ and $\lambda = 0.011(2)~\mu $s$^{-1}$ at 0.05~K\@.  The fits are shown by solid red curve in Fig.~\ref{fig:musr1}. Since within the error bars both $\sigma$ and $\lambda$ at $T<T_{\rm c}$ and $T>T_{\rm c}$ are similar, there is no evidence of time reversal symmetry breaking in ${\rm LaIrSi_3}$.

\begin{figure}
\includegraphics[width=\columnwidth, keepaspectratio]{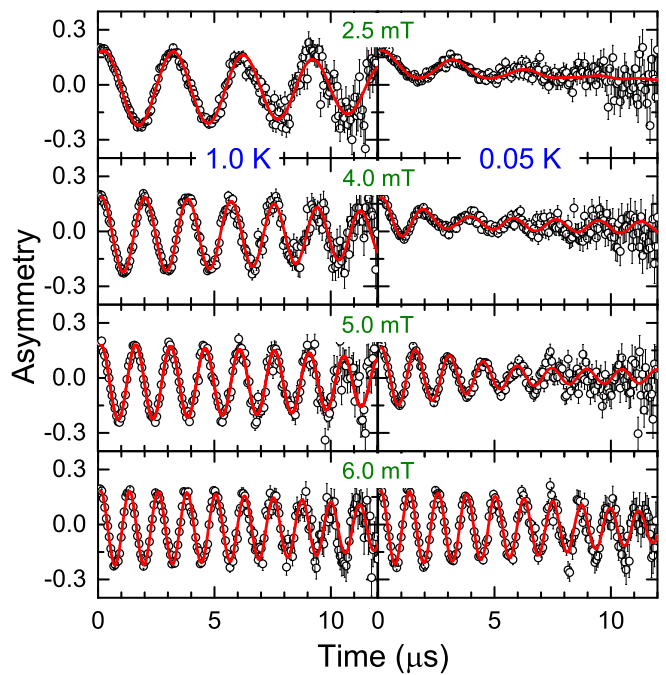}
\caption{\label{fig:musr2} (colour online) The transverse field $\mu$SR spectra at temperatures above (1.0~K, left panels) and below (0.05~K, right panels) the superconducting $T_{\rm c}$ for indicated applied fields. The solid curves are the fits by oscillatory function in Eq.~(\ref{eq:TF}).}
\end{figure}

\begin{figure}
\includegraphics[width=3in, keepaspectratio]{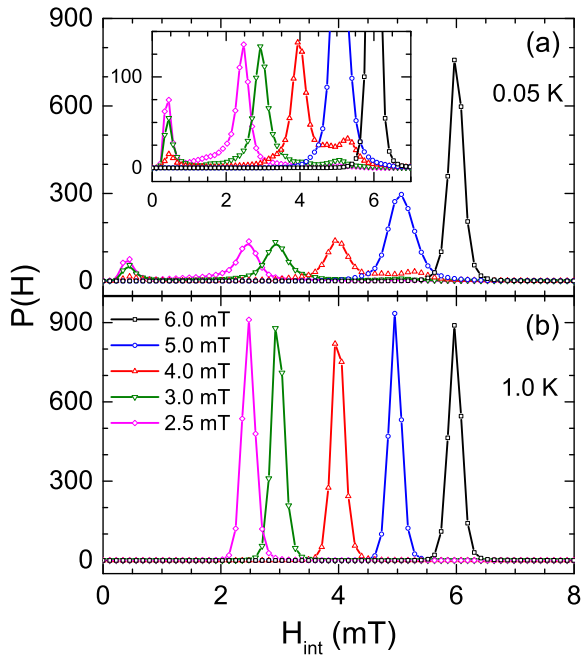}
\caption{\label{fig:musr3} (colour online) The maximum entropy spectra at different applied fields $H$ for (a) 0.05~K, and (b) 1.0~K\@. Inset in (a) shows an expanded view of the magnetic field probability distribution $P(H)$.}
\end{figure}

The time evolution of muon spin rotation in transverse field (TF) is shown in Fig.~\ref{fig:musr2}. The TF muon spin precession signals were collected on field cooled sample at different applied fields both above (1.0~K) and below (0.05~K) $T_{\rm c}$. The TF $\mu$SR spectra are best described by an oscillatory function damped with a Gaussian and an oscillatory background, i.e. by
\begin{equation}
\label{eq:TF}
 Gz(t)= A_0 \cos\left(\omega t + \varphi\right) {\rm e}^{ -\sigma^2 t^2/2} + A_{\rm BG} \cos\left(\omega t + \varphi\right)
\end{equation}
where $\omega = \gamma_{\mu}H_{\rm int}$ is the precession frequency ($H_{\rm int}$ is the internal field at muon site). Solid curves in Fig.~\ref{fig:musr2} are the fits of the TF $\mu$SR spectra by the decay function in Eq.~(\ref{eq:TF}). In the superconducting state at $T =0.05$~K the depolarization rate is found to increase significantly, e.g. for the TF $\mu$SR spectra at 2.5~mT the $\sigma$ increases from its value $\sigma = 0.010(2)~\mu $s$^{-1}$ at 1.0~K to $\sigma = 1.45(4)~\mu$s$^{-1}$ at 0.05~K\@. Such an increase of depolarization rate reveals bulk superconductivity in ${\rm LaIrSi_3}$.

The maximum entropy spectra for TF $\mu$SR precession at 1.0~K and 0.05~K are shown in Fig.~\ref{fig:musr3}. The maximum entropy spectra depicts the magnetic field probability distribution $P(H)$. It is clear from Fig.~\ref{fig:musr3} that at 1.0~K (in normal state) sharp peaks are observed at $H_{\rm int}$ exactly equal to the applied $H$, whereas at 0.05~K (in superconducting state) one can see additional peaks. At $H =4.0$~mT the $P(H)$ at 0.05~K shows an additional peak at $H_{\rm int} > H$ [inset of Fig.~\ref{fig:musr3}]. The appearance of an additional peak (near 5.5~mT, which gives an estimation of $H_{c}$) at an internal field greater than the applied $H$ is a characteristic of a type-I behavior and indicates a type-I superconductivity in ${\rm LaIrSi_3}$ consistent with the above inference from the value of $\kappa_{\rm GL}$ in Table~\ref{tab:SCParams}. Further at 0.05~K for $H \leq 4.0$~mT we also observe an increase in $P(H)$ near $H_{\rm int} \sim 0.5$~mT as is expected for a sample in Meissner state. However, no such increase in $P(H)$ is observed from Meissner volume in the intermediate state (i.e. at $H = 5.0$~mT). At $H =6.0$~mT no additional peak is observed in $P(H)$ at $H_{\rm int} > H$ even on expanded scale. This could be understood to be due to the fact that the applied $H$ is close to the $H_{c} = 6.1$~mT (see Table~\ref{tab:SCParams}) and the sample is on the verge of transition from superconducting to normal state at $H =6.0$~mT. Thus the $\mu$SR data also reflect a low thermodynamic critical field $H_{c}\approx 5.5$~mT in line with the estimate of $H_{c}$ from the heat capacity data.

\section{\label{Conclusion} Conclusions}

The superconducting and normal state properties of noncentrosymmetric superconductor LaIrSi$_{3}$ which crystallizes in BaNiSn$_{3}$-type tetragonal crystal structure (space group $I4\,mm$) are investigated by $C_{\rm p}(T,H)$, $\rho(T,H)$ and $\mu$SR measurements which demonstrate bulk BCS superconductivity below $T_{\rm c} = 0.72(1)$~K\@. A nonbulk superconductivity sets in at a higher $T_{\rm c}$ in $\rho (T)$. In the normal state the $\rho$ exhibits metallic behavior and $\rho(T\geq1.6~{\rm K})$  is well described by the Bloch-Gr\"{u}neisen model of resistivity. Our analysis of low temperature normal state $C_{\rm p}(T)$ data yield Sommerfeld coefficient $\gamma_{\rm n} = 4.60(2) $~mJ/mol\,K$^{2}$ corresponding to the density of states at Fermi energy ${\cal D}(E_{\rm F}) = 1.95(1)$~states/eV\,f.u.\ for both spin directions.

The superconducting transition is revealed by a very sharp jump in $C_{\rm p}$ at $T_{\rm c} = 0.72(1)$~K however with a reduced value of $\Delta C_{\rm e}/\gamma_{\rm n} T_{\rm c}  =1.09(3)$ than the BCS expected value of 1.43. The reduced value of $\Delta C_{\rm e}/\gamma_{\rm n} T_{\rm c}$ seems to indicate an anisotropic energy gap in ${\rm LaIrSi_3}$. The superconducting state electronic heat capacity data are analyzed by single-band $\alpha$-model of BCS superconductivity that describes the experimental data reasonably. The $\alpha = \Delta(0)/k_{\rm B}T_{\rm c} = 1.54(2)$ obtained from the jump in $C_{\rm p}$ is smaller than the $\alpha_{\rm BCS} = 1.764$ indicating the $s$-wave order parameter of ${\rm LaIrSi_3}$ to be anisotropic in momentum space. Even though the single-band $\alpha$-model describes the superconducting state data reasonably, considering the split spin-up and spin-down bands in ${\rm LaIrSi_3}$ the possibility of two-band superconductivity cannot be ruled out. Presence of two-band is also known to result in a reduced $\Delta C_{\rm e}/\gamma_{\rm n} T_{\rm c}$. \cite{Johnston2013,Zehetmayer2013} Since the two-band effect and anisotropy modification to single-band have similar manifestations it is very difficult to distinguish between them by examining the thermodynamic quantities. Recently a two-band superconductivity with equal energy gaps was reported in SrPt$_3$P. \cite{Khasanov2014} Further investigations are desired to check for the possibility of a similar two-band single-gap superconductivity in ${\rm LaIrSi_3}$.

Various normal and superconducting state parameters have been estimated which indicate a dirty limit weak-coupling type-I $s$-wave BCS superconductivity in ${\rm LaIrSi_3}$. Type-I superconductivity is further confirmed by $\mu$SR. The $\mu$SR measurement also revealed that the time-reversal symmetry is preserved in superconducting state thus confirming a conventional $s$-wave singlet pairing superconductivity in ${\rm LaIrSi_3}$. Thus despite a large splitting of Fermi surfaces due to antisymmetric coupling on account of noncentrosymmetric structure inferred from de Haas-van Alphen effect study \cite{Okuda2007} no clear signature of parity mixing or spin-triplet pairing state is found in our investigations of superconducting state properties of ${\rm LaIrSi_3}$.

The type-I superconductivity in LaIrSi$_3$ is similar to that of NCS LaRhSi$_3$ which also exhibits conventional $s$-wave electron-phonon mediated type-I superconductivity with preserved time reversal symmetry and a singlet pairing. \cite{Anand2011a}  Both LaRhSi$_3$ and LaIrSi$_3$ have similar ${\cal D}(E_{\rm F})$ and $\Theta_{\rm D}$ therefore one would expect the $T_{\rm c}$ of these two compounds to be similar, however we see that the $T_{\rm c} = 0.72(1)$~K of LaIrSi$_3$ is significantly lower than the $T_{\rm c} = 2.16(8)$~K of LaRhSi$_3$.\cite{Anand2011a} While the reason for this contrasting behavior is not clear, the $T_{\rm c}$ appears to have a relation with the outer shell electronic configuration of Rh ($4d$) and Ir ($5d$) in these two superconductors. A similar trend has been observed in other $4d$ and $5d$ systems such as in LaPdSi$_3$ [$T_{\rm c} =2.65(5)$~K] and LaPtSi$_3$ [$T_{\rm c} =1.52(6)$~K] (Ref.~\onlinecite{Smidman2014}) as well as in Li$_2$Pd$_3$B [$T_{\rm c}=6.7$~K] and Li$_2$Pt$_3$B [$T_{\rm c}=2.43$~K]. \cite{Yuan2006} The $4d$-based superconductors appears to have higher $T_{\rm c}$ than those having $5d$. Further investigations to understand this aspect of superconductivity in these compounds are desired.

The effect of lack of inversion symmetry on the superconducting properties of LaIrSi$_3$ seems not pronounced despite a large splitting of spin-up and spin-down energy bands due to ASOC. A similar behavior is reported for NCS BaPtSi$_{3}$ (Ref.~\onlinecite{Bauer2009}) for which electronic-structure calculations revealed splitting of bands on account of spin-orbit interactions, however the superconductivity turned out to be conventional BCS-like with a singlet paring. Such observations raise an important question: what else other than ASOC controls the appearance of anomalous superconducting state in a noncentrosymmetric system? The anomalous superconducting properties of the Ce-based strongly correlated NCSs, such as CePt$_{3}$Si, \cite{Bauer2007} CeRhSi$_{3}$ (Ref.~\onlinecite{Kimura2007}) and CeIrSi$_{3}$ (Ref.~\onlinecite{Sugitani2006}) can naively be related to the magnetic pairing due to the presence of $4f$ moments. However, the unusual superconducting properties of weakly correlated NCSs Li$_2$Pt$_3$B,\cite{Yuan2006} LaNiC$_{2}$, \cite{Hillier2009,Bonalde2011} and Re$_6$Zr (Ref.~\onlinecite{Singh2014}) are not then obvious as they do not show any evidence of magnetic order. An important difference between the two groups of nonmagnetic NCSs, i.e. those exhibiting unusual superconducting properties and those exhibiting conventional superconductivity is the difference in their crystal structures/space groups. This may suggest that these two groups of nonmagnetic NCSs may have different Fermi surface topology that may have some role in realizing the effect of ASOC. A comparative study of extent of ASOC in nonmagnetic NCSs preferably by the technique that can directly probe Fermi surface topology, such as angle resolved photoemission spectroscopy (ARPES), complemented with band-structure calculations can shed light on this issue and would be of help in understanding the relationship of ASOC, Fermi surface topology and anomalous superconductivity in NCSs.

\acknowledgments

VKA, DTA and ADH acknowledge financial assistance from CMPC-STFC grant number CMPC-09108. AB thanks UJ and STFC for PDF funding. AMS thanks the SA-NRF (78832) and the URC of UJ for financial assistance.

\end{document}